\documentclass[10pt,twocolumn]{IEEEtran}
\usepackage[latin1]{inputenc}
\usepackage{amssymb}
\usepackage{latexsym}
\usepackage{mathrsfs}
\usepackage{amsfonts}
\usepackage{amsbsy}
\usepackage{amsmath}
\usepackage{times}
\usepackage{epsfig,epsf,times,amssymb,amsmath}
\usepackage{color}
\usepackage{setspace}

\title{Distributed Space-Time Coding Based on Adjustable Code Matrices for Cooperative MIMO Relaying Systems}

\author{Tong Peng, Rodrigo C. de Lamare,~\IEEEmembership{Senior Member,~IEEE,}
and Anke Schmeink, ~\IEEEmembership{Member,~IEEE}
\thanks{Tong Peng, Rodrigo C. de Lamare is with Communications Research Group, Department of Electronics, University of York, York YO10 5DD, UK, e-mails: tp525@ohm.york.ac.uk; rcdl500@ohm.york.ac.uk}
\thanks{Anke Schmeink, UMIC Research Centre, RWTH Aachen University, D-52056 Aachen, Germany, e-mail: schmeink@umic.rwth-aachen.de}
\thanks{Part of this work has been presented at ISWCS 2012.}}

\begin{document}
\maketitle

\begin{abstract}
An adaptive distributed space-time coding (DSTC) scheme
is proposed for two-hop cooperative MIMO networks. Linear minimum
mean square error (MMSE) receive filters and adjustable code matrices are considered subject to a power constraint with an amplify-and-forward (AF) cooperation strategy. In the proposed adaptive DSTC scheme, an adjustable code matrix obtained by a feedback channel is employed to transform the space-time coded matrix at the relay node. The effects of the limited feedback and the feedback errors are
assessed. Linear MMSE expressions are devised to compute the
parameters of the adjustable code matrix and the linear receive
filters. Stochastic gradient (SG) and least-squares (LS) algorithms are also developed with
reduced computational complexity. An upper bound on the pairwise error probability analysis is derived and indicates the advantage of employing the adjustable code matrices at the relay nodes. An alternative optimization algorithm for the adaptive DSTC scheme is also derived in order to eliminate the need for the feedback. The algorithm provides a fully distributed scheme for the adaptive DSTC at the relay node based on the minimization of the error probability. Simulation results show that the proposed algorithms obtain significant performance gains as compared to existing DSTC schemes.
\end{abstract}

\begin{IEEEkeywords}
Adaptive algorithms, space-time codes with feedback, cooperative systems, distributed space time codes.
\end{IEEEkeywords}

\section{Introduction}

Cooperative multiple-input and multiple-output (MIMO) systems, which
employ multiple relay nodes with antennas between the source node
and the destination node as a distributed antenna array, can obtain
diversity gains by providing copies of the transmitted signals to
improve the reliability of wireless communication systems
\cite{Clarke}. Among the links between the relay nodes and the
destination node, cooperation strategies such as Amplify-and-Forward
(AF), Decode-and-Forward (DF), Compress-and-Forward (CF) \cite{J. N.
Laneman2004} and various distributed space-time coding (DSTC)
schemes \cite{J. N. Laneman2003}, \cite{Yiu S.}, \cite{RC De Lamare}
can be employed.

The use of distributed space-time codes (DSTC) at the relay node in
a cooperative network, providing more copies of the desired symbols
at the destination node, can offer the system diversity and coding
gains to mitigate the interference. A recent focus on DSTC
techniques lies on the design of delay-tolerant codes and
full-diversity schemes with minimum outage probability. An
opportunistic DSTC scheme with the minimum outage probability is
designed for a DF cooperative network and compared with the fixed
DSTC schemes in \cite{Yulong}, while in \cite{Behrouz Maham} a novel
opportunistic relaying algorithm is achieved by employing DSTC in an
AF cooperative MIMO network. An adaptive distributed-Alamouti
(D-Alamouti) space-time block code (STBC) design is proposed in
\cite{Abouei} for non-regenerative dual-hop wireless systems which
achieves the minimum outage probability. DSTC schemes for the AF
protocol are discussed in \cite{Maham}-\cite{Maham Birsen}. In
\cite{Maham}, the GABBA STC scheme is extended to a distributed MIMO
network with full-diversity and full-rate, while an optimal
algorithm for the design of the DSTC scheme to achieve the optimal
diversity and multiplexing tradeoff is derived in \cite{Sheng}. A
quasi-orthogonal DSTBC for cooperative MIMO networks is presented
and shown to achieve full rate and full diversity with any number of
antennas in \cite{Maham Birsen}. In \cite{Birsen Sirkeci-Mergen}, a
new STC scheme that multiplies a randomized matrix by the STC matrix
at the relay node before the transmission is derived and analyzed.
The randomized space-time code (RSTC) can achieve the performance of
a centralized space-time code in terms of coding gain and diversity
order.

Optimal space-time codes can be obtained by transmitting the channel
or other useful information for code design back to the source node,
in order to achieve higher coding gains by pre-processing the
symbols. In \cite{Mari}, the trade-off between the length of the
feedback symbols, which is related to the capacity loss and the
transmission rate is discussed, whereas in \cite{Amir} one solution
for this trade-off problem is derived. The use of limited feedback
for STC encoding has been widely discussed in the literature. In
\cite{Jabran}, the phase information is sent back for STC encoding
in order to maintain the full diversity, and the phase feedback is
employed in \cite{IIhwan} to improve the performance of the Alamouti
STBC. A limited feedback link is used in \cite{George} and
\cite{David} to provide the channel information for the pre-coding
of an orthogonal STBC scheme. Another limited feedback strategy has
been considered for power relay selection in \cite{TDS_2}.

In this paper, we propose an adaptive distributed space-time coding
scheme and algorithms for cooperative MIMO relaying systems. This
work was first introduced and discussed in \cite{Tong}. We first
develop a centralized algorithm with limited feedback to compute the
parameters of an adjustable code matrix, which requires sending the
adjustable code matrices back to the relay nodes after the
optimization via a feedback channel that is modeled as a Binary
Symmetric Channel (BSC). Then, adaptive optimization algorithms are
derived based on the MSE and the ML criteria subject to constraints
on the transmitted power at the relays, in order to release the
destination node from the high computational complexity of the
optimization process. We focus on how the adjustable code matrix
affects the DSTC during the encoding and how to optimize the linear
receive filter with the code matrix iteratively or, alternatively,
by employing an ML detector and adjusting the code matrix. The upper
bound of the error probability of the proposed adaptive DSTC is
derived in order to show its advantages as compared to the
traditional DSTC schemes and the influence of the imperfect feedback
is discussed. It is shown that the use of an adjustable code matrix
benefits the performance of the system compared to employing
traditional STC schemes. Then, we derive a fully distributed matrix
optimization algorithm which does not require feedback. The pairwise
error probability (PEP) of the adaptive DSTC is employed in order to
devise a distributed algorithm and to eliminate the need for
feedback channels. The fully distributed matrix optimization
algorithm allows the system to use the optimal adjustable matrix
before the transmission, and also achieves the minimum PEP when the
statistical information of the channel does not change. The
differences of our work compared with the existing works are
discussed as follows. First, an optimal adjustable code matrix will
be multiplied by an existing space-time coding scheme at the relay
node and the encoded data are forwarded to the destination node. The
code matrix is first generated randomly as discussed in \cite{Birsen
Sirkeci-Mergen}, and it is optimized according to different criteria
at the destination node by the proposed algorithms. Second, in order
to implement the adaptive algorithms, the adjustable code matrix is
optimized with the linear receive filter iteratively, and then
transmitted back to the relay node via a feedback channel. The
impact of the feedback errors is considered and shown in the
simulations. Third, the proposed fully distributed optimization
algorithm eliminates the effect of the feedback by choosing the
optimal code matrix before transmission, and the receiver is
released from the design task.

The paper is organized as follows. Section II introduces a two-hop
cooperative MIMO system with multiple relays applying the AF
strategy and the adaptive DSTC scheme. In Section III the proposed
optimization algorithms for the adjustable code matrix are derived, and the pairwise error probability is analyzed in Section IV. The fully distributed optimization algorithm is derived in Section V, and the results of the simulations are given in Section VI. Section VII gives the conclusions of the work.

Notation: the italic, bold lower-case and bold upper-case letters denote scalars, vectors and matrices, respectively. The operators $E[\cdot]$ and $(\cdot)^\emph{H}$ stand for expected value and the Hermitian operator. The $N \times N$ identity matrix is written as ${\boldsymbol I}_N$. $\parallel{\boldsymbol X}\parallel_F=\sqrt{{\rm Tr}({\boldsymbol X}^H\cdot{\boldsymbol X})}=\sqrt{{\rm Tr}({\boldsymbol X}\cdot{\boldsymbol X}^\emph{H})}$ is the Frobenius norm. $\Re[\cdot]$ and $\Im[\cdot]$ stand for the real part and the imaginary part, respectively. $Tr(\cdot)$ stands for the trace of a matrix, and $(\cdot)^{\dag}$ for pseudo-inverse, and $\bigotimes$ denotes the Kronecker product.

\section{Cooperative MIMO System Model}

\begin{figure}
\begin{center}
\def\epsfsize#1#2{0.825\columnwidth}
\epsfbox{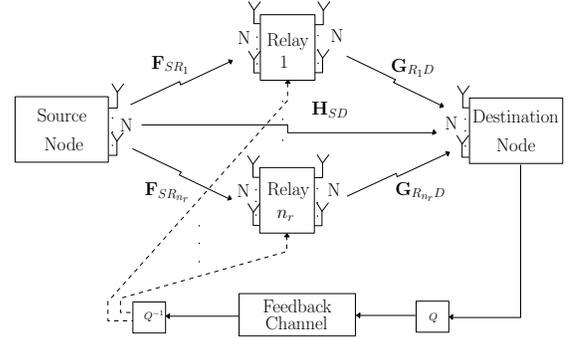}\vspace*{-1em} \caption{Cooperative MIMO
system model with $n_r$ relay nodes}\label{1}
\vspace{-3em}
\end{center}
\end{figure}

The communication system under consideration is a two-hop cooperative MIMO system which employs multiple relay nodes as shown in Fig. 1. The first hop is devoted to the source transmission, which broadcasts the information symbols to the relay nodes and to the destination node. The second hop forwards the amplified and re-encoded information symbols from the relay nodes to the destination node. An orthogonal transmission protocol is considered which requires that the source node does not transmit during the time period of the second hop. In order to evaluate the adaptive optimization algorithms, a BSC is considered as the feedback channel.

Consider a cooperative MIMO system with $n_r$ relay nodes that employ an AF cooperative strategy as well as a DSTC scheme. All nodes have $N$ antennas to transmit and receive. We consider only one user at the source node in our system that operates in a spatial multiplexing configuration. Let ${\boldsymbol s}[i]$ denotes the transmitted information symbol vector at the source node, which contains $N$ parameters, ${\boldsymbol s}[i] = [s_{1}[i], s_{2}[i], ... , s_{N}[i]]$, and has a covariance matrix $E\big[ {\boldsymbol s}[i]{\boldsymbol s}^{H}[i]\big] = \sigma_{s}^{2}{\boldsymbol I}_N$, where $\sigma_s^2$ is the signal power which we assume to be equal to 1. The source node broadcasts ${\boldsymbol s}[i]$ from the source to $n_r$ relay nodes as well as to the destination node in the first hop, which can be described by
\begin{equation}\label{2.1}
{\boldsymbol r}_{SD}[i] = {\boldsymbol H}_{SD}[i]{\boldsymbol s}[i] + {\boldsymbol n}_{SD}[i],~~~
{\boldsymbol r}_{{SR}_{k}}[i]  = {\boldsymbol F}_{SR_{k}}[i]{\boldsymbol s}[i] + {\boldsymbol n}_{SR_k}[i],
\end{equation}
\begin{equation*}
i = 1,2,~...~,N, ~~k = 1,2,~...~ n_{r},
\end{equation*}
where ${\boldsymbol r}_{{SR}_{k}}[i]$ and ${\boldsymbol r}_{SD}[i]$ denote
the received symbol vectors at the $k$th relay node and at the destination node,
respectively. The $N \times 1$ vector ${\boldsymbol n}_{{SR}_{k}}[i]$ and ${\boldsymbol n}_{SD}[i]$ denote the zero mean complex circular symmetric additive white Gaussian noise (AWGN) vector generated at the $k$th relay node and at the destination node with variance $\sigma^{2}$. The matrices ${\boldsymbol F}_{SR_k}[i]$ and ${\boldsymbol H}_{SD}[i]$ are the $N \times N$ channel coefficient matrices between the source node and the $k$th relay node, and between the source node and the destination node, respectively.

The received symbols are amplified and re-encoded at each relay node prior to transmission to the destination node in the second hop. We assume that the synchronization at each node is perfect. After amplifying the received vector ${\boldsymbol r}_{SR_k}[i]$ at the $k$th relay node, the signal vector $\tilde{{\boldsymbol s}}_{SR_k}[i]={\boldsymbol A}_{R_kD}[i]{\boldsymbol r}_{SR_k}[i]$ can be obtained, where ${\boldsymbol A}_{R_kD}[i]$ stands for the $N \times N$ diagonal amplification matrix assigned at the $k$th relay node. The $N \times 1$ signal vector $\tilde{{\boldsymbol s}}_{SR_k}[i]$ will be re-encoded by an $N \times T$ DSTC scheme ${\boldsymbol M}(\tilde{{\boldsymbol s}})$, multiplied by an $N \times N$ adjustable code matrix ${\boldsymbol \Phi}_k[i]$ generated randomly \cite{Birsen Sirkeci-Mergen}, and then forwarded to the destination node. The relationship between the $k$th relay and the destination node can be described as
\begin{equation}\label{2.2}
{\boldsymbol R}_{R_{k}D}[i] = {\boldsymbol G}_{R_kD}[i]{\boldsymbol \Phi}_k[i]{\boldsymbol M}_{R_{k}D}[i] + {\boldsymbol N}_{R_{k}D}[i].
\end{equation}
The $N \times T$ received symbol matrix ${\boldsymbol R}_{R_{k}D}[i]$ in
(\ref{2.2}) can be written as an $NT \times 1$ vector ${\boldsymbol
r}_{R_{k}D}[i]$ given by
\begin{equation}\label{2.3}
{\boldsymbol r}_{R_{k}D}[i]  = {\boldsymbol \Phi}_{eq_k}[i]{\boldsymbol
G}_{{eq}_k}[i]\tilde{{\boldsymbol s}}_{SR_k}[i] + {\boldsymbol
n}_{R_{k}D}[i],
\end{equation}
where the block diagonal $NT \times NT$ matrix ${\boldsymbol
\Phi}_{eq_k}[i]$ denotes the equivalent adjustable code matrix
and the $NT \times N$ matrix ${\boldsymbol G}_{eq_k}[i]$ stands for
the equivalent channel matrix which is the DSTC scheme ${\boldsymbol
M}(\tilde{{\boldsymbol s}}[i])$ combined with the channel matrix
${\boldsymbol G}_{R_{k}D}[i]$. The $NT \times 1$ equivalent noise
vector ${\boldsymbol n}_{R_{k}D}[i]$ generated at the destination
node contains the noise parameters in ${\boldsymbol N}_{R_{k}D}[i]$.

The use of an adjustable code matrix or a randomized matrix
${\boldsymbol \Phi}_{eq_k}[i]$ which achieves the full diversity
order and provides a lower error probability has been discussed in
\cite{Birsen Sirkeci-Mergen}. The uniform sphere randomized matrix
which achieves the lowest BER of the analyzed schemes and contains
elements that are uniformly distributed on the surface of a complex
hyper-sphere of radius $\rho$ is used in our system. The proposed
adaptive algorithms detailed in the next section optimize the code
matrices employed at the relay nodes in order to achieve a lower
BER. At each relay node, the adjustable code matrices are normalized
so that no increase in the energy is introduced at the relay nodes
and the comparison between different schemes is fair.

After rewriting ${\boldsymbol R}_{R_{k}D}[i]$ we can consider the
received symbol vector at the destination node as a $(T+1)N \times 1$
vector with two parts, one is from the source node and another one
is the superposition of the received vectors from each relay node.
Therefore, the received symbol vector for the cooperative MIMO
system can be written as
\begin{equation}\label{2.4}
\begin{aligned}
{\boldsymbol r}[i]  &=
\left[\begin{array}{c} {\boldsymbol H}_{SD}[i]{\boldsymbol s}[i]  \\ \sum_{k=1}^{n_r}{\boldsymbol \Phi}_{eq_k}[i]{\boldsymbol G}_{{eq}_k}[i]\tilde{{\boldsymbol s}}_{SR_k}[i] \end{array} \right] + \left[\begin{array}{c}{\boldsymbol n}_{SD}[i] \\ {\boldsymbol n}_{RD}[i] \end{array} \right] \\
& = {\boldsymbol D}_D[i]\tilde{\boldsymbol s}_D[i] + {\boldsymbol n}_D[i],
\end{aligned}
\end{equation}
where the $(T + 1)N \times 2N$ block diagonal matrix
${\boldsymbol D}_D[i]$ denotes the channel gain matrix of all the
links in the network which contains the $N \times N$ channel
coefficients matrix ${\boldsymbol H}_{SD}[i]$ between the source node and
the destination node, the $NT \times N$ equivalent channel matrix
${\boldsymbol G}_{{eq}_k}[i]$ for $k=1,2,...,n_r$ between each relay
node and the destination node. We assume that the coefficients in all channel matrices are independent and remain constant over the transmission. The $(T + 1)N \times 1$ noise vector ${\boldsymbol n}_D[i]$ contains the equivalent received noise vector at the destination node, which can be modeled as an AWGN with zero mean and covariance matrix $\sigma^{2}(1+\parallel\sum_{k=1}^{n_r}{\boldsymbol \Phi}_{eq_k}[i]{\boldsymbol G}_{{eq}_k}[i]{\boldsymbol
A}_{R_kD}[i]\parallel^2_F){\boldsymbol I}_{(T+1)N}$.

\section{Joint Adaptive Code Matrix Optimization and Receiver Design}

In this section, we jointly design an MMSE adjustable code matrix
and the receiver for the proposed DSTC scheme. Adaptive SG and RLS
algorithms \cite{S. Haykin} for determining the parameters of the
adjustable code matrix with reduced complexity are also devised. The
DSTC scheme used at the relay node employs an MMSE-based adjustable
code matrix, which is computed at the destination node and obtained
by a feedback channel in order to process the data symbols prior to
transmission to the destination node. It is worth to mention that
the code matrices are only used at the relay node so the direct link
from the source node to the destination node is not considered in
this section.

\subsection{Linear MMSE Receiver Design with Adaptive DSTC Optimization}

The linear MMSE receiver design with optimal code matrices is
derived as follows. By defining the $TN \times 1$ parameter vector
${\boldsymbol w}_j[i]$ to determine the $j$th symbol $s_j[i]$, we
propose the MSE based optimization with a power constraint at the
destination node described by
\begin{equation}\label{3.1}
    [{\boldsymbol w}_j[i],{\boldsymbol \Phi}_{eq_k}[i]] = \arg\min_{{\boldsymbol w}_j[i], {\boldsymbol \Phi}_{eq_k}[i]} E\left[\|s_j[i]-{\boldsymbol w}_j^\emph{H}[i]{\boldsymbol r}[i]\|^2\right], ~s.t.~
    \rm{Tr}({\boldsymbol \Phi}_{eq_k}[i]{\boldsymbol \Phi}_{eq_k}^\emph{H}[i])\leq \rm{P_R},
\end{equation}
where ${\boldsymbol r}[i]$ denotes the received symbol vector at the
destination node. By employing a Lagrange multiplier $\lambda$ we
can obtain the Lagrange expression shown as
\begin{equation}\label{3.1.1}
    \mathscr{L}=E\left[\|s_j[i]-{\boldsymbol w}_j^\emph{H}[i]{\boldsymbol r}[i]\|^2\right]+\lambda(\rm{Tr}({\boldsymbol \Phi}_{eq_k}[i]{\boldsymbol \Phi}_{eq_k}^\emph{H}[i])-\rm{P_R}).
\end{equation}

By expanding the right-hand side of (\ref{3.1.1}) and taking the
gradient with respect to ${\boldsymbol w}_j^*[i]$ and equating the
terms to zero, we can obtain the $j$th MMSE receive filter vector
for the $j$th symbol
\begin{equation}\label{3.2}
    {\boldsymbol w}_j[i]={\boldsymbol R}^{-1}{\boldsymbol p},
\end{equation}
where the first term ${\boldsymbol R}=E\left[{\boldsymbol
r}[i]{\boldsymbol r}^\emph{H}[i]\right]$ denotes the
auto-correlation matrix and the second term ${\boldsymbol
p}=E\left[{\boldsymbol r}[i]s_j^*[i]\right]$ stands for the
cross-correlation vector. To optimize the code matrix ${\boldsymbol
\Phi}_{{eq_k}_j}[i]$ for each symbol at each relay node, we can
calculate the code matrix by taking the gradient with respect to
${\boldsymbol \Phi}_{{eq_k}_j}^*[i]$ and equating the terms to zero,
resulting in
\begin{equation}\label{3.3}
    {\boldsymbol \Phi}_{{eq_k}_j}[i]= \tilde{\boldsymbol R}^{-1}\tilde{\boldsymbol P},
\end{equation}
where $\tilde{\boldsymbol
R}=E\left[s_j[i]\tilde{s}_{SR_{k_j}}[i]{\boldsymbol
w}_j[i]{\boldsymbol w}^\emph{H}_j[i]+\lambda{\boldsymbol I}\right]$
and $\tilde{\boldsymbol
P}=E\left[s_j[i]\tilde{s}_{SR_{k_j}}[i]{\boldsymbol
w}_j[i]{\boldsymbol g}^\emph{H}_{eq_{k_j}}[i]\right]$ are $NT \times
NT$ matrices. The value of the Lagrange multiplier $\lambda$ can be
determined by substituting ${\boldsymbol \Phi}_{{eq_k}_j}[i]$ into
$\lambda\rm{Tr}({\boldsymbol \Phi}_{eq_k}[i]{\boldsymbol
\Phi}_{eq_k}^\emph{H}[i])= \rm{P_R}$ and solving the power
constraint function. In the proposed adaptive algorithm we employ
quantization instead of using the Lagrange multiplier, which
requires less computational complexity. The detailed explanation is
shown in the next section. Note that non-linear detection algorithms
\cite{delamare_spa} can also be employed at the receiver for an
improved performance.

Appendix A includes a detailed derivation of ${\boldsymbol w}_j[i]$
and ${\boldsymbol \Phi}_{eq_j}[i]$. The power constraint can be
enforced by employing the Lagrange multiplier and by substituting
the power constraint into the MSE cost function. In (\ref{3.3}) a
closed-form expression of the code matrix ${\boldsymbol
\Phi}_{{eq_k}_j}[i]$ assigned for the $j$th received symbol at the
$k$th relay node is derived. The problem is that the optimization
method requires the calculation of a matrix inversion with a high
computational complexity of ${\rm O}((NT)^3)$, and with the increase
in the number of antennas employed at each node or the use of more
complicated STC encoders at the relay nodes, the computational
complexity increases cubically according to the matrix sizes in
(\ref{3.2}) and (\ref{3.3}).

\subsection{Adaptive Stochastic Gradient Optimization Algorithm}

In order to reduce the computational complexity and achieve an
optimal performance, a centralized adaptive robust matrix
optimization (C-ARMO) algorithm based on an SG algorithm with a
linear receiver design is proposed as follows.

The Lagrangian resulting from the optimization problem is derived in
(\ref{3.1.1}), and a simple adaptive algorithm for determining the
linear receive filters and the code matrices can be derived by
taking the instantaneous gradient term of (\ref{3.1.1}) with respect
to ${\boldsymbol w}^*_j[i]$ and with respect to ${{\boldsymbol
\Phi}^*_{{eq_k}_j}}[i]$, respectively, which are
\begin{equation}\label{3.4}
\begin{aligned}
     \nabla \mathscr{L}_{{\boldsymbol w}_j^*[i]} & = \nabla E\left[\|s_j[i]-{\boldsymbol w}_j^\emph{H}[i]{\boldsymbol r}[i]\|^2\right]_{{\boldsymbol w}_j^*[i]} = -e_j^*[i]{\boldsymbol r}[i],\\
     \nabla \mathscr{L}_{{\boldsymbol \Phi}_{{eq_k}_j}^*[i]} &= \nabla E\left[\|s_j[i]-{\boldsymbol w}_j^\emph{H}[i]{\boldsymbol r}[i]\|^2\right]_{{{\boldsymbol \Phi}^*_{{eq_k}_j}}[i]} = -e_j[i]s^*_j[i]{\boldsymbol w}_j[i]{\boldsymbol d}_{k_j}^\emph{H}[i],
\end{aligned}
\end{equation}
where $e_j[i]=s_j[i]-{\boldsymbol w}_j^\emph{H}[i]{\boldsymbol r}[i]$ stands for the $j$th error signal, and the $NT \times 1$ vector ${\boldsymbol d}_{k_j}[i]$ denotes the $j$th column of the channel matrix which contains the product of the channel matrices ${\boldsymbol F}_{SR_k}$ and ${\boldsymbol G}_{R_kD}$ and the power allocation matrices ${\boldsymbol A}_{R_kD}$. After we obtain (\ref{3.4}) the proposed algorithm can be obtained by introducing a step size into a gradient optimization algorithm to update the result until the convergence is reached, and the algorithm is given by
\begin{equation}\label{3.5}
     {\boldsymbol w}_j[i+1] = {\boldsymbol w}_j[i] + \beta (e_j^*[i]{\boldsymbol r}[i]), ~~~
     {\boldsymbol \Phi}_{{eq_k}_j}[i+1] = {\boldsymbol \Phi}_{{eq_k}_j}[i] +\mu (e_j[i]s^*_j[i]{\boldsymbol w}_j[i]{\boldsymbol d}_{k_j}^\emph{H}[i]),
\end{equation}
where $\beta$ and $\mu$ denote the step sizes in the recursions for the estimation. A detailed derivation is included in Appendix B.

The energy of the code matrices in (\ref{3.5}) will be increased with the processing of the adaptive algorithm, which will contribute to the reduction of the error probability. A normalization of the code matrix after the optimization is required and implemented as ${\boldsymbol \Phi}_{{eq_k}_j}[i+1] =\frac{\sqrt{\rm{P_R}}{\boldsymbol \Phi}_{{eq_k}_j}[i+1]}{\sqrt{\sum_{j=1}^{N}\rm{Tr}({\boldsymbol \Phi}_{{eq_k}_j}[i+1]{\boldsymbol \Phi}_{{eq_k}_j}^\emph{H}[i+1])}}$ to ensure that the energy is not increased and for a fair comparison among the analyzed DSTC schemes. A summary of the C-ARMO SG algorithm is given in Table I.
\begin{table}
  \centering
  \caption{Summary of the C-ARMO SG Algorithm}\label{}
  \begin{tabular}{cc}
  \hline
1: & Initialize: ${\boldsymbol w}_j[0] = {\boldsymbol 0}_{NT \times 1}$, \\
2: & ${\boldsymbol \Phi}[0]$ is generated randomly with the power constraint $\rm{Tr}({\boldsymbol \Phi}_{eq_k}{\boldsymbol \Phi}_{eq_k}^\emph{H})\leq \rm{P_R}$. \\
3: & For each instant of time, $i$=1, 2, ..., compute \\
4: & $\nabla \mathscr{L}_{{\boldsymbol w}_j^*[i]} = -e_j^*[i]{\boldsymbol r}[i]$, \\
5: & $\nabla \mathscr{L}_{{\boldsymbol \Phi}_{{eq_k}_j}^*[i]} = -e_j[i]s^*_j[i]{\boldsymbol w}_j[i]{\boldsymbol d}_{k_j}^\emph{H}[i]$, \\
6: & where $e_j[i]=s_j[i]-{\boldsymbol w}_j^\emph{H}[i]{\boldsymbol r}[i]$. \\
7: & Update ${\boldsymbol w}_j[i]$ and ${\boldsymbol \Phi}_{{eq_k}_j}[i]$ by \\
8: & ${\boldsymbol w}_j[i+1] = {\boldsymbol w}_j[i] + \beta (e_j^*[i]{\boldsymbol r}[i])$,\\
9: & ${\boldsymbol \Phi}_{{eq_k}_j}[i+1] ={\boldsymbol \Phi}_{{eq_k}_j}[i] +\mu (e_j[i]s^*_j[i]{\boldsymbol w}_j[i]{\boldsymbol d}_{k_j}^\emph{H}[i])$,\\
10: & ${\boldsymbol \Phi}_{{eq_k}_j}[i+1] = \frac{\sqrt{\rm{P_R}}{\boldsymbol \Phi}_{{eq_k}_j}[i+1]}{\sqrt{\sum_{j=1}^{N}\rm{Tr}({\boldsymbol \Phi}_{{eq_k}_j}[i+1]{\boldsymbol \Phi}_{{eq_k}_j}^\emph{H}[i+1])}}$.\\
\hline
\end{tabular}
\vspace{-2em}
\end{table}

According to (\ref{3.5}), the receive filter ${\boldsymbol w}_j[i]$
and the code matrix ${\boldsymbol \Phi}_{{eq_k}_j}[i]$ depend on
each other. Therefore, alternating optimization algorithms
\cite{jidf,jiomimo} can be used to determine the linear MMSE receive
filter and the code matrix iteratively, and the optimization
procedure can be completed. The complexity of calculating the
optimal ${\boldsymbol w}_j[i]$ and ${\boldsymbol
\Phi}_{{eq_k}_j}[i]$ is ${\rm O}(NT)$ and ${\rm O}(N^2T^2)$,
respectively, which is much less than $O(N^4T^4)$ and $O(N^5T^5)$ by
using (\ref{3.2}) and (\ref{3.3}). As mentioned in Section I, the
optimal MMSE code matrices will be sent back to the relay nodes via
a feedback channel, and the influence of the imperfect feedback is
shown and discussed in simulations.

\subsection{ML Detection and LS Code Matrix Estimation Algorithm}

The criterion for optimizing the adjustable code matrices and
performing symbol detection in the C-ARMO algorithm can be changed
to the maximum likelihood (ML) criterion, which is equivalent to a
Least-squares (LS) criterion in this case. For example, if we take
the ML instead of the MSE criterion to determine the code matrices,
then we have to store an $N \times D$ matrix ${{\boldsymbol S}}$ at
the destination node which contains all the possible combinations of
the transmitted symbol vectors. The ML optimization problem can be
written as
\begin{equation}\label{3.7}
    [\hat s_{d_j}[i],\hat {\boldsymbol \Phi}_{eq_{k_j}}[i]]= \arg\min_{s_{d_j}[i],{\boldsymbol \Phi}_{eq_{k_j}}[i]} \|{\boldsymbol r}[i]-{\boldsymbol {\hat{r}}}[i]\|^2,
    ~s.t.~ \rm{Tr}({\boldsymbol \Phi}_{eq_k}[i]{\boldsymbol \Phi}_{eq_k}^\emph{H}[i])\leq \rm{P_R}, ~for~ d = 1,2,...,D,
\end{equation}
where ${\boldsymbol {\hat{r}}}[i]=\sum_{k=1}^{n_r}\sum_{j=1}^{N}{\boldsymbol \Phi}_{eq_{k_j}}[i]{\boldsymbol d}_{k_j}[i]\hat s_{d_j}[i]$ denotes the received symbol vector without noise which is determined by substituting each column of ${\boldsymbol S}$ into (\ref{3.7}). It is worth to mention that the optimization algorithm contains a discrete part which refers to the ML detection and a continuous part which refers to the optimization of the code matrix, and the detection and the optimization can be implemented separately as they do not depend on each other. The optimization algorithm can be considered as a mixed discrete-continues optimization. In this case, other detectors such as sphere decoders can be used in the optimization algorithm in the detection part in order to reduce the computational complexity without an impact on the performance, and the algorithm will converge after several iterations.

After determining the transmitted symbol vector, we can calculate the optimal code matrix ${\boldsymbol \Phi}_{{eq_k}_j}[i]$ by employing the LS estimation algorithm. The Lagrangian expression is given by
\begin{equation}\label{3.8}
    \mathscr{L}=\|{\boldsymbol r}[i]-(\sum_{k=1}^{n_r}\sum_{j=1}^{N}{\boldsymbol \Phi}_{eq_{k_j}}[i]{\boldsymbol d}_{k_j}[i]\hat s_{d_j}[i])\|^2+\lambda(Tr[{\boldsymbol \Phi}_{eq_k}[i]{\boldsymbol \Phi}^\emph{H}_{eq_k}[i]]-P_R),
\end{equation}
and by taking the instantaneous gradient of $\mathscr{L}$ with respect to the code matrix ${\boldsymbol \Phi}^*_{eq_{k_j}}[i]$ we can obtain
\begin{equation}\label{3.9}
\begin{aligned}
    \nabla \mathscr{L}_{{\boldsymbol \Phi}^*_{eq_{k_j}}[i]} & = ({\boldsymbol r}[i]-{\boldsymbol {\hat{r}}}[i])\nabla_{{\boldsymbol \Phi}^*_{eq_{k_j}}[i]}({\boldsymbol r}[i]-{\boldsymbol {\hat{r}}}[i])^\emph{H}\\
    & = ({\boldsymbol r}_{e_j}[i]-{\boldsymbol \Phi}_{eq_{k_j}}[i]{\boldsymbol d}_{k_j}[i]\hat s_{d_j}[i])(-\hat s^*_{d_j}[i]{\boldsymbol d}\emph{H}_{k_j}[i]),
\end{aligned}
\end{equation}
where ${\boldsymbol r}_{e_j}[i]={\boldsymbol r}[i]-\sum_{k=1}^{n_r}\sum_{l=1,l\neq j}^{N}{\boldsymbol \Phi}_{eq_{k_l}}[i]{\boldsymbol d}_{k_l}[i]\hat{s}_{d_l}[i]$ stands for the received vector without the desired code matrix. The optimal code matrix $\hat {\boldsymbol \Phi}_{eq_{k_j}}[i]$ requires $\nabla \mathscr{L}_{{\boldsymbol \Phi}_{{eq_k}_j}^*[i]}=0$, and the optimal adjustable code matrix as given by
\begin{equation}\label{3.10}
    {\boldsymbol \Phi}_{{eq_k}_j}[i]=\hat{s}_{d_j}^*[i]{\boldsymbol r}_{e_j}[i]{\boldsymbol d}^\emph{H}_{k_j}[i](\mid \hat{s}_{d_j}[i]\mid^2{\boldsymbol d}_{k_j}[i]{\boldsymbol d}^\emph{H}_{k_j}[i])^{\dag}.
\end{equation}
The power constraint is not considered because the quantization
method can be employed in order to reduce the high computational
complexity for determining the value of the Lagrange multiplier.

\subsection{RLS Code Matrix Estimation Algorithm}
The RLS estimation algorithm for the code matrix ${\boldsymbol \Phi}_{{eq_k}_j}[i]$ is derived in this section. The ML detector is employed so that the detection and the optimization procedures are separate as explained in the previous section, so we focus on how to optimize the code matrix rather than the detection. The superior convergence behavior of the LS algorithm when the size of the adjustable code matrix is large indicates the reason of the utilization of an RLS estimation, and it is worth to mention that the computational complexity reduces from cubic to square by employing the RLS algorithm.

According to the RLS algorithm, the optimization problem is given by
\begin{equation}\label{3.11}
    [\hat {\boldsymbol \Phi}_{eq_{k_j}}[i]]= \arg\min_{{\boldsymbol \Phi}_{eq_{k_j}}[i]} \sum_{n=1}^{i}\lambda^{i-n}\|{\boldsymbol r}[n]-{\boldsymbol {\hat{r}}}[i]\|^2,
    ~s.t.~ \rm{Tr}({\boldsymbol \Phi}_{eq_k}[i]{\boldsymbol \Phi}_{eq_k}^\emph{H}[i])\leq \rm{P_R},
\end{equation}
where $\lambda$ stands for the forgetting factor. By expanding the right-hand side of (\ref{3.11}) and taking gradient with respect to ${\boldsymbol \Phi}^*_{eq_{k_j}}[i]$ and equaling the terms to zero, we obtain
\begin{equation}\label{3.12}
    {\boldsymbol \Phi}_{eq_{k_j}}[i]=(\sum_{n=1}^{i}\lambda^{i-n}{\boldsymbol r}_e[n]{\boldsymbol r}_{k_j}^\emph{H}[n])(\sum_{i=1}^{n}\lambda^{i-n}{\boldsymbol r}_{k_j}[n]{\boldsymbol r}_{k_j}^\emph{H}[n])^{-1},
\end{equation}
where the $NT \times 1$ vector ${\boldsymbol r}_e[n]={\boldsymbol \Phi}_{eq_{k_j}}[n]{\boldsymbol d}_{k_j}[n]\hat s_{d_j}[n]$ and ${\boldsymbol r}_{k_j}[n]={\boldsymbol d}_{k_j}[n]\hat s_{d_j}[n]$. The power constraint is still not considered during the optimization. We define
\begin{equation}\label{3.13}
    {\boldsymbol \Psi}[i] = \sum_{n=1}^{i}\lambda^{i-n}{\boldsymbol r}_{k_j}[n]{\boldsymbol r}_{k_j}^\emph{H}[n] = \lambda {\boldsymbol \Psi}[i-1]+{\boldsymbol r}_{k_j}[n]{\boldsymbol r}_{k_j}^\emph{H}[n],
\end{equation}
\begin{equation}\label{3.14}
    {\boldsymbol Z}[i] = \sum_{n=1}^{i}\lambda^{i-n}{\boldsymbol r}_e[n]{\boldsymbol r}_{k_j}^\emph{H}[n]=\lambda {\boldsymbol Z}[i-1]+{\boldsymbol r}_e[n]{\boldsymbol r}_{k_j}^\emph{H}[n],
\end{equation}
so that we can rewrite (\ref{3.12}) as
\begin{equation}\label{3.15}
    {\boldsymbol \Phi}_{eq_{k_j}}[i] = {\boldsymbol Z}[i]{\boldsymbol \Psi}^{-1}[i].
\end{equation}
By employing the matrix inversion lemma in \cite{Tylavsky}, we can obtain
\begin{equation}\label{3.16}
    {\boldsymbol \Psi}^{-1}[i]=\lambda^{-1}{\boldsymbol \Psi}^{-1}[i-1]-\lambda^{-1}{\boldsymbol k}[i]{\boldsymbol r}_{k_j}^\emph{H}[i]{\boldsymbol \Psi}^{-1}[i-1],
\end{equation}
where ${\boldsymbol k}[i]=(\lambda^{-1}{\boldsymbol \Psi}^{-1}[i-1]{\boldsymbol r}_{k_j}[i])/(1+\lambda^{-1}{\boldsymbol r}_{k_j}^H[i]{\boldsymbol \Psi}^{-1}[i-1]{\boldsymbol r}_{k_j}[i])$. We define ${\boldsymbol P}[i]={\boldsymbol \Psi}^{-1}[i]$ and by substituting (\ref{3.14}) and (\ref{3.16}) into (\ref{3.15}), the expression of the code matrix is given by
\begin{equation}\label{3.17}
\begin{aligned}
    {\boldsymbol \Phi}_{eq_{k_j}}[i] &= \lambda{\boldsymbol Z}[i-1]{\boldsymbol P}[i]+{\boldsymbol r}_e[i]{\boldsymbol r}_{k_j}^\emph{H}[i]{\boldsymbol P}[i]\\ &= {\boldsymbol Z}[i-1]{\boldsymbol P}[i-1]+{\boldsymbol Z}[i-1]{\boldsymbol k}[i]{\boldsymbol r}_{k_j}^\emph{H}[i]{\boldsymbol P}[i-1]+{\boldsymbol r}_e[i]{\boldsymbol r}_{k_j}^\emph{H}[i]{\boldsymbol P}[i]\\ &= {\boldsymbol \Phi}_{eq_{k_j}}[i-1]+\lambda^{-1}({\boldsymbol r}_e[i]-{\boldsymbol Z}[i-1]{\boldsymbol k}[i]){\boldsymbol r}_{k_j}^\emph{H}[i]{\boldsymbol P}[i-1].
\end{aligned}
\end{equation}

Table II shows a summary of the C-ARMO RLS algorithm.
\begin{table}
  \centering
  \caption{Summary of the C-ARMO RLS Algorithm}\label{}
  \begin{tabular}{cc}
  \hline
1: & Initialize: ${\boldsymbol P}[0]=\delta^{-1}{\boldsymbol I}_{NT \times NT}$, ${\boldsymbol Z}[0] = {\boldsymbol I}_{NT \times NT}$, \\
2: & the value of $\delta$ is small when SNR is high and is large when SNR is low,\\
3: & ${\boldsymbol \Phi}[0]$ is generated randomly with the power constraint $\rm{trace}({\boldsymbol \Phi}_{eq_k}[i]{\boldsymbol \Phi}_{eq_k}^\emph{H}[i])\leq \rm{P_R}$. \\
4: & For each instant of time, $i$=1, 2, ..., compute \\
5: & ${\boldsymbol k}[i]=\frac{\lambda^{-1}{\boldsymbol \Psi}^{-1}[i-1]{\boldsymbol r}_{k_j}[i]}{1+\lambda^{-1}{\boldsymbol r}_{k_j}^\emph{H}[i]{\boldsymbol \Psi}^{-1}[i-1]{\boldsymbol r}_{k_j}[i]}$,\\
6: & ${\boldsymbol \Phi}_{eq_{k_j}}[i]  = {\boldsymbol \Phi}_{eq_{k_j}}[i-1]+\lambda^{-1}({\boldsymbol r}_e[i]-{\boldsymbol Z}[i-1]{\boldsymbol k}[i]){\boldsymbol r}_{k_j}^\emph{H}[i]{\boldsymbol P}[i-1]$, \\
7: & ${\boldsymbol P}[i]=\lambda^{-1}{\boldsymbol P}[i-1]-\lambda^{-1}{\boldsymbol k}[i]{\boldsymbol r}^\emph{H}_{k_l}[i]{\boldsymbol P}[i-1]$, \\
8: & ${\boldsymbol Z}[i] = \lambda {\boldsymbol Z}[i-1]+{\boldsymbol r}_e[i]{\boldsymbol r}_{k_j}^\emph{H}[i]$. \\
12: & ${\boldsymbol \Phi}_{{eq_k}_j}[i] = \frac{\sqrt{\rm{P_R}}{\boldsymbol \Phi}_{{eq_k}_j}[i]}{\sqrt{\sum_{j=1}^{N}\rm{Tr}({\boldsymbol \Phi}_{{eq_k}_j}[i]{\boldsymbol \Phi}_{{eq_k}_j}^\emph{H}[i])}}$.\\
\hline
\end{tabular}
\vspace{-2em}
\end{table}

\subsection{Convergence Analysis}
The C-ARMO algorithms can be divided into two cases: the first one performs the optimization by updating the receive filter and the code matrix iteratively, i.e., the MSE based C-ARMO algorithm, the second one only optimizes the code matrix itself according to the Lagrangian function, i.e., the ML and RLS based C-ARMO algorithms. In this subsection, we will illustrate how the C-ARMO algorithms converge to the global optimum solution.

\subsubsection{MSE based C-ARMO algorithm}
The proposed MSE based C-ARMO algorithm allows the optimization of the receive filter ${\boldsymbol w}[i]$ and the code matrix ${\boldsymbol \Phi}[i]$ iteratively. A detailed proof of the convergence of this type of algorithm is derived in \cite{UNiesen}. We will give a brief outline on how these results can be used to prove the convergence of our algorithms.

According to \cite{UNiesen}, the optimization problem in (\ref{3.1}) can be described as: Given an initial $({\boldsymbol w}_0,{\boldsymbol \Phi}_0)\in \mathcal {W}_0 \times \mathcal{P}_0$, we have to find a sequence of points $({\boldsymbol w}_n,{\boldsymbol \Phi}_n)\in \mathcal {W}_n \times \mathcal{P}_n$ that
\begin{equation}\label{3.18}
    \lim_{n\rightarrow\infty}\mathscr{L}({\boldsymbol w}_n,{\boldsymbol \Phi}_n)=\mathscr{L}(\mathcal {W}, \mathcal{P}),
\end{equation}
where the sequence of compact sets $\{(\mathcal {W}_n, \mathcal{P}_n)\}_{n \geq 0} : \mathcal {W}_n, \mathcal{P}_n$ that are revealed at time $n$ such that as $n\rightarrow\infty$, $\mathcal{P}_n\stackrel{d_H}{\rightarrow}\mathcal{P}$ and $\mathcal{W}_n\stackrel{d_H}{\rightarrow}\mathcal{W}$, and
\begin{equation}\label{3.19}
    d_H(\mathcal {A},\mathcal {B})=\max\{\sup_{A\in\mathcal {A}}\inf_{B\in\mathcal {B}}d(A,B),\sup_{B\in\mathcal {B}}\inf_{A\in\mathcal {A}}d(A,B)\}
\end{equation}
denotes the Hausdorff distance between $\mathcal {A}$ and $\mathcal {B}$. The proposed algorithm is written recursively for $n\geq 1$ as described by
\begin{equation}\label{3.20}
    {\boldsymbol w}_n \in \arg\min_{{\boldsymbol w}\in\mathcal {W}_n}\mathscr{L}({\boldsymbol w},{\boldsymbol \Phi}_{n-1}),~ {\boldsymbol \Phi}_n \in \arg\min_{{\boldsymbol \Phi}\in\mathcal {P}_n}\mathscr{L}({\boldsymbol w}_n,{\boldsymbol \Phi}).
\end{equation}
According to the three-point property and the four-point property in \cite{UNiesen}, we can obtain
\begin{equation}\label{3.21}
    \mathscr{L}({\boldsymbol w}_n,{\boldsymbol \Phi}_n)+\mathscr{L}({\boldsymbol w},{\boldsymbol \Phi}_n) \leq \mathscr{L}({\boldsymbol w},{\boldsymbol \Phi}_{n-1})+\mathscr{L}({\boldsymbol w},{\boldsymbol \Phi})+\omega(\gamma_n)
\end{equation}
for all ${\boldsymbol w} \in \mathcal {W}_n$ and ${\boldsymbol \Phi} \in \mathcal {P}_n$, where $\omega(\gamma_n)$ denotes the modulus of continuity of $\mathscr{L}({\boldsymbol w}_n,{\boldsymbol \Phi}_n)$ with $\omega(\gamma_n) \rightarrow 0$ as $\gamma_n \rightarrow 0$, and $\gamma_n=\varepsilon_n+\varepsilon_{n-1}$ with $\varepsilon_n \rightarrow 0$ as $n \rightarrow \infty$.

Since $\mathcal{P}_n\stackrel{d_H}{\rightarrow}\mathcal{P}$ and $\mathcal{W}_n\stackrel{d_H}{\rightarrow}\mathcal{W}$, there must exists a sequence $({\boldsymbol w}'_n,{\boldsymbol \Phi}'_n) \in \mathcal {W}_n \times \mathcal {P}_n$ such that $({\boldsymbol w}'_n,{\boldsymbol \Phi}'_n) \rightarrow ({\boldsymbol w}',{\boldsymbol \Phi}') \in \arg\min \mathscr{L}(\mathcal {W},\mathcal {P})$ and $d({\boldsymbol w}'_n,{\boldsymbol w}')+d({\boldsymbol \Phi}'_n,{\boldsymbol \Phi}') \leq \varepsilon_n$ for all $n \geq 0$. By replacing $({\boldsymbol w},{\boldsymbol \Phi})$ with $({\boldsymbol w}',{\boldsymbol \Phi}')$ and choosing $\mathscr{L}({\boldsymbol w}'_n,{\boldsymbol \Phi}'_n) \leq \mathscr{L}({\boldsymbol w}',{\boldsymbol \Phi}')+\omega(\varepsilon_n)$, we can obtain
\begin{equation}\label{3.22}
    \mathscr{L}({\boldsymbol w}_n,{\boldsymbol \Phi}_n)+\mathscr{L}({\boldsymbol w}'_n,{\boldsymbol \Phi}_n) \leq \mathscr{L}({\boldsymbol w}'_{n-1},{\boldsymbol \Phi}_{n-1})+\mathscr{L}({\boldsymbol w}',{\boldsymbol \Phi}')+2\omega(\gamma_n)+\omega(\varepsilon_n),
\end{equation}
and further derive
\begin{equation}\label{3.23}
    \liminf_{n \rightarrow \infty}\mathscr{L}({\boldsymbol w}_n,{\boldsymbol \Phi}_n) \leq \mathscr{L}({\boldsymbol w},{\boldsymbol \Phi}).
\end{equation}

By defining a subsequence $\{n_k\}_{k>0}$ such that $\liminf_{n \rightarrow \infty}\mathscr{L}({\boldsymbol w}_n,{\boldsymbol \Phi}_n)=\lim_{k \rightarrow \infty}\mathscr{L}({\boldsymbol w}_{n_k},{\boldsymbol \Phi}_{n_k})$, and assuming compactness of $\mathcal {W}$ and $\mathcal {P}$, we can obtain ${\boldsymbol w} \in \mathscr{W}$, ${\boldsymbol \Phi} \in \mathcal {P}$, and
\begin{equation}\label{3.24}
    \liminf_{n \rightarrow \infty}\mathscr{L}({\boldsymbol w}_n,{\boldsymbol \Phi}_n) = \lim_{k \rightarrow \infty}\mathscr{L}({\boldsymbol w}_{n_k},{\boldsymbol \Phi}_{n_k}) \geq \mathscr{L}({\boldsymbol w},{\boldsymbol \Phi}).
\end{equation}
Combining (\ref{3.23}) and (\ref{3.24}), we can obtain $\liminf_{n \rightarrow \infty}\mathscr{L}({\boldsymbol w}_n,{\boldsymbol \Phi}_n) =\mathscr{L}({\boldsymbol w},{\boldsymbol \Phi})$ which indicates (\ref{3.18}) converges to the optimum values.

\subsubsection{ML and RLS based C-ARMO algorithm}
The ML and RLS based C-ARMO algorithms just optimize the code matrix,  and we can analyze the Hessian matrix of (\ref{3.8}) and check its positive (semi-)definiteness. By taking the second-order partial derivatives of the Lagrangian cost function in (\ref{3.8}), we can obtain
\begin{equation}\label{3.25}
\begin{aligned}
    H(\mathscr{L}) &= \frac{\partial}{\partial{\boldsymbol \Phi}_{eq_{k_j}}[i]}(\frac{\partial \mathscr{L}}{\partial {\boldsymbol \Phi}^*_{eq_{k_j}}[i]}) = \frac{\partial}{\partial{\boldsymbol \Phi}_{eq_{k_j}}[i]}({\boldsymbol r}[i]s^*_j[i]{\boldsymbol d}^\emph{H}_{k_j}[i]+\mid s_j\mid^2{\boldsymbol \Phi}_{eq_{k_j}}[i]{\boldsymbol d}_{k_j}[i]{\boldsymbol d}^\emph{H}_{k_j}[i])\\ &=\mid s_j\mid^2{\boldsymbol d}_{k_j}[i]{\boldsymbol d}^\emph{H}_{k_j}[i],
\end{aligned}
\end{equation}
where the first term $\mid s_j\mid^2$ is a positive scalar and the rest of the terms denotes the multiplication of the equivalent channel vectors which is a positive-definite matrix and the problem is convex. We conclude that the Hessian matrix of the Lagrangian cost function is a positive-definite matrix so that the ML and RLS based C-ARMO algorithms converge to the global optimum under the usual assumptions used to prove the convergence of these algorithms for convex problems.

\section{Probability of Error Analysis}

In this section, the pairwise error probability (PEP) of the system employing the adaptive DSTC will be derived. As we mentioned in Section I, the adjustable code matrices will be considered in the derivation as it affects the performance by reducing the upper bound of the pairwise error probability. The PEP upper bound of the traditional STC schemes in \cite{Hamid} is introduced for comparison, and the main difference lies in the eigenvalues of the adjustable code matrices. Please note that the direct link is ignored in the PEP upper bound derivation in order to concentrate on the effects of the adjustable code matrix on the performance. The expression of the upper bound holds for systems with different sizes and an arbitrary number of relay nodes.

Consider an $N \times N$ STC scheme at the relay node with
$T$ codewords, and the codeword ${\boldsymbol C}^1$ is transmitted and
decoded as another codeword ${\boldsymbol C}^i$ at the destination
node, where $i=1,2,...,T$. According to \cite{Hamid}, the
probability of error for this code can be upper bounded by the sum of all the
probabilities of incorrect decoding, which is given by
\begin{equation}\label{4.1.1}
    {\rm P_e} \leq \sum_{i=2}^{T} {\rm P}({\boldsymbol C}^1\rightarrow{\boldsymbol C}^i).
\end{equation}
Assuming that the codeword ${\boldsymbol C}^2$ is decoded at the destination node and that we know the channel information perfectly, we can derive the conditional pairwise error probability of the DSTC encoded with the adjustable code matrix ${\boldsymbol \Phi}$ as \cite{JYuan}
\begin{equation}\label{4.1}
    {\rm P}({\boldsymbol C}^1\rightarrow{\boldsymbol C}^2\mid{{\boldsymbol \Phi}})  = {\rm Q}\left(\sqrt{\frac{\gamma}{2}}\parallel{\boldsymbol \Phi}{\boldsymbol D}({\boldsymbol C}^1-{\boldsymbol C}^2)\parallel_F\right),
\end{equation}
where ${\boldsymbol D}$ stands for the matrix with the channel coefficients for all links. Let ${\boldsymbol U}^\emph{H}{\boldsymbol \Lambda}_{\boldsymbol C}{\boldsymbol U}$ be the eigenvalue decomposition of $({\boldsymbol C}^1-{\boldsymbol C}^2)^\emph{H}({\boldsymbol C}^1-{\boldsymbol C}^2)$, where ${\boldsymbol U}$ is a unitary matrix with the eigenvectors and ${\boldsymbol \Lambda}_{\boldsymbol C}$ is a diagonal matrix which contains all the eigenvalues of the difference between two different codewords ${\boldsymbol C}^1$ and ${\boldsymbol C}^2$. Let ${\boldsymbol V}^\emph{H}{\boldsymbol \Lambda}_{\boldsymbol \Phi}{\boldsymbol V}$ stand for the eigenvalue decomposition of $({\boldsymbol \Phi}{\boldsymbol D}{\boldsymbol U})^H{\boldsymbol \Phi}{\boldsymbol D}{\boldsymbol U}$, where ${\boldsymbol V}$ is a unitary matrix that contains the eigenvectors and ${\boldsymbol \Lambda}_{\boldsymbol \Phi}$ is a diagonal matrix with the eigenvalues arranged in decreasing order. Therefore, the conditional pairwise probability of error can be written as
\begin{equation}\label{4.4}
    {\rm P}({\boldsymbol C}^1\rightarrow{\boldsymbol C}^2\mid{\boldsymbol \Phi})={\rm Q}\left(\sqrt{\frac{\gamma}{2}\sum^{NT}_{m=1}\sum^N_{n=1}\lambda_{{\boldsymbol \Phi}_n}\lambda_{{\boldsymbol C}_n}|\xi_{n,m}|^2}\right),
\end{equation}
where $\xi_{n,m}$ is the $(n,m)$th element in ${\boldsymbol V}$, and $\lambda_{{\boldsymbol \Phi}_n}$ and $\lambda_{{\boldsymbol C}_n}$ are the $n$th eigenvalues in ${\boldsymbol \Lambda}_{\boldsymbol \Phi}$ and ${\boldsymbol \Lambda}_s$, respectively. It is important to note that the value of $\lambda_{\boldsymbol \Phi}$ is positive and real because $({\boldsymbol \Phi}{\boldsymbol D}{\boldsymbol U})^H{\boldsymbol \Phi}{\boldsymbol D}{\boldsymbol U}$ is Hermitian symmetric. According to \cite{Hamid}, an appropriate upper bound assumption of the ${\rm Q}$ function is ${\rm Q}(x)\leq\frac{1}{2}e^{\frac{-x^2}{2}}$, thus we can derive the upper bound of the pairwise error probability for an adaptive STC scheme as
\begin{equation}\label{4.5}
    {\rm P}_{e_{\boldsymbol \Phi}}\leq E \left[\frac{1}{2}\exp\left(-\frac{\gamma}{4}\sum^{NT}_{m=1}\sum^N_{n=1}\lambda_{{\boldsymbol \Phi}_n}\lambda_{{\boldsymbol C}_n}|\xi_{n,m}|^2\right) \right]=\frac{1}{\prod_{n=1}^{N}(1+\frac{\gamma}{4}\lambda_{{\boldsymbol \Phi}_n}\lambda_{{\boldsymbol C}_n})^{NT}},
\end{equation}
while the upper bound of the error probability expression for a traditional STC in \cite{Hamid} is given by
\begin{equation}\label{4.6}
    {\rm P}_{e_{\boldsymbol D}}\leq E \left[\frac{1}{2}\exp\left(-\frac{\gamma}{4}\sum^{NT}_{m=1}\sum^N_{n=1}\lambda_{{\boldsymbol C}_n}|\xi_{n,m}|^2\right)\right]=\frac{1}{\prod_{n=1}^{N}(1+
    \frac{\gamma}{4}\lambda_{{\boldsymbol C}_n})^{NT}}.
\end{equation}
If we neglect the $1$ in the denominator in (\ref{4.5}), the exponent of the SNR $\gamma$ indicates the diversity order which means the full diversity $NTN$ can be achieved in (\ref{4.5}). By comparing (\ref{4.5}) and (\ref{4.6}), employing an adjustable code matrix for an STC scheme at the relay node introduces $\lambda_{{\boldsymbol \Phi}_n}$ in the BER upper bound. With the aid of simulations, we found that ${\boldsymbol \Lambda}_{\boldsymbol \Phi}$ is diagonal with one eigenvalue less than $1$ and others much greater than $1$. As a result, employing the adjustable code matrices can provide a decrease in the BER upper bound since the value in the denominator increases.

\section{The fully distributed adaptive robust matrix optimization algorithm}
Inspired by the analysis developed in the previous section, we derive a fully distributed ARMO (FD-ARMO) algorithm which does not require the feedback channel in this section. We will extend the exact PEP expression in \cite{Giorgio} for MIMO communication systems to the AF cooperative MIMO systems with the adaptive DSTC schemes. Then, we design the FD-ARMO algorithm to determine and store the adjustable code matrices at the relay nodes before the transmission in Phase II.

The exact PEP expression of an STC has been given by Taricco and Biglieri in \cite{Giorgio}, which contains the sum of the real part and the imaginary part of the mean value of the error probability, and the moment generating function (MGF) is employed to compute the mean value. To extend the exact PEP expression to the cooperative MIMO systems, we can rewrite the received symbol vector at the destination node as
\begin{equation*}
    {\boldsymbol R}_{RD} = \sum_{k=1}^{n_r}{\boldsymbol \Phi}_k[i]{\boldsymbol D}_k[i]{\boldsymbol C}[i]+{\boldsymbol N}_{RD}[i],
\end{equation*}
where ${\boldsymbol D}_k[i]$ denotes the channel matrix. For simplicity, we assume the synchronization is perfect, and each relay node transmits the STC matrix simultaneously and the received symbol vector at the destination node will be the superposition of each column of each STC code. The equivalent noise vector contains the AWGN at the destination node as well as the amplified and re-encoded noise vectors at the relay nodes. As a result the PEP expression of the AF cooperative MIMO system with the adaptive DSTC can be derived as
\begin{equation}\label{5.1}
    {\rm P}({\boldsymbol C}^1\rightarrow{\boldsymbol C}^2\mid{\boldsymbol \Phi}_{eq})={\rm Q}\left(\frac{\parallel{\boldsymbol \Phi}{\boldsymbol D}({\boldsymbol C}^1-{\boldsymbol C}^2)\parallel_F}{\sqrt{2N_o}}\right),
\end{equation}
where $N_0={\rm Tr}({\boldsymbol I}+{\boldsymbol \Phi}{\boldsymbol D})$ denotes the received noise variance at the destination node. We define ${\boldsymbol \Delta} = {\boldsymbol C}^1-{\boldsymbol C}^2$ as the distance between the code words, and $\tau = \sqrt{\frac{1}{2N_o}}{\boldsymbol \Phi}{\boldsymbol D}{\boldsymbol \Delta}{\boldsymbol \Delta}^\emph{H}{\boldsymbol D}^\emph{H}{\boldsymbol \Phi}^\emph{H}$ and we assume that the eigenvalue decomposition of ${\boldsymbol \Delta}{\boldsymbol \Delta}^\emph{H}$ can be written as ${\boldsymbol V}{\boldsymbol \Lambda}{\boldsymbol V}^\emph{H}$, where ${\boldsymbol V}$ stands for a unitary matrix that contains the eigenvectors of ${\boldsymbol \Delta}{\boldsymbol \Delta}^H$ and ${\boldsymbol \Lambda}$ contains all the eigenvalues of the square of the distance vector. Define an $N \times N$ matrix ${\boldsymbol Z}={\boldsymbol \Phi}{\boldsymbol D}$, and ${\boldsymbol Z}\sim \textit{N}_c(\mu_{\boldsymbol Z},{\boldsymbol \Sigma}_{\boldsymbol Z})$, where $\mu_{\boldsymbol Z}=0$ denotes the mean and ${\boldsymbol \Sigma}_{\boldsymbol Z}=E\left[{\boldsymbol \Sigma}_{\boldsymbol Z}{\boldsymbol \Sigma}_{\boldsymbol Z}^\emph{H}\right]$ stands for the covariance matrix. The expression of the error probability is given by
\begin{equation}\label{5.2}
\begin{aligned}
    \Theta(c) & = E\left[ \exp(-c\xi)\right]=E\left[ \exp(-c\sqrt{\frac{1}{2N_o}}[{\boldsymbol \Phi}{\boldsymbol D}\Delta\Delta^\emph{H}{\boldsymbol D}^\emph{H}{\boldsymbol \Phi}^\emph{H}]) \right]\\
    &=E\left[ \exp\left(-c\sqrt{\frac{1}{2N_o}}[{\boldsymbol Z}{\boldsymbol \Lambda}{\boldsymbol Z}^\emph{H}]\right) \right] = \frac{\exp\left(-\mu_{\boldsymbol Z}^\emph{H}{\boldsymbol B}({\boldsymbol I}+\Sigma_{\boldsymbol Z}{\boldsymbol B})^{-1}\mu_{\boldsymbol Z}\right)}{\det\left({\boldsymbol I} + \frac{c}{2 \sqrt{2N_0}}{\boldsymbol \Phi}{\boldsymbol \Lambda}{\boldsymbol \Phi}^\emph{H} \right)}\\
    & = \det\left({\boldsymbol I} + \frac{c}{2 \sqrt{2N_0}}{\boldsymbol \Phi}{\boldsymbol \Lambda}{\boldsymbol \Phi}^\emph{H} \right)^{-1},
\end{aligned}
\end{equation}
where ${\boldsymbol B}={\boldsymbol I}\bigotimes {\boldsymbol \Delta}{\boldsymbol \Delta}^\emph{H}$, and $c=a+jb$ is the variable defined in the MGF with $a=\frac{1}{4}$ and $b$ is a constant. By inserting (\ref{5.2}) into the pairwise error probability expression in \cite{Giorgio}, we can obtain the exact PEP of the adaptive DSTC scheme written as
\begin{equation}\label{5.3}
    {\rm P}_e = \frac{1}{2J}\sum_{i=1}^{J}\{\Re[\Phi(c)]+\frac{b}{a}\Im[\Phi(c)]\}+{\rm E}_J,
\end{equation}
where ${\rm E}_J \rightarrow 0$ as $J \rightarrow \infty$.

Since the PEP is proportional to (\ref{5.2}), it is clear that minimizing the PEP is equal to maximizing the determinant of ${\boldsymbol I} + \frac{c}{2 \sqrt{2N_o}}{\boldsymbol \Phi}{\boldsymbol \Lambda}{\boldsymbol \Phi}^\emph{H}$. As a result, the optimization problem can be written as
\begin{equation}\label{5.4}
    \Theta_{opt}(c)=\arg\max_l \Theta_l(c),~~~~l=1,2,...
\end{equation}
where $\Theta_l(c)$ stands for the $l$th candidate code matrix. For simplicity the candidate code matrices are generated randomly and satisfy the power constraint. In order to obtain the adjustable code matrix we can first randomly generate a set of matrices, and then substitute them into (\ref{5.2}) to compute the determinant. In the simulation, we randomly generate $500$ code matrices and choose the optimal one according to the FD-ARMO algorithm. The optimal code matrix with the largest value of the determinant which achieves the minimal PEP will be employed at the relay node. A summary of the FD-ARMO is given in Table III. It is worth to mention that the FD-ARMO algorithm is non-convex, but it can still achieve the optimal performance by choosing the optimal code matrix from a number of candidates even though this not guaranteed.

\begin{table}
  \centering
  \caption{Summary of the FD-ARMO Algorithm}\label{}
  \begin{tabular}{cc}
  \hline
1: & Choose the $N \times T$ STC scheme used at the relay node\\
2: & Determine the dimension of the adjustable code matrix ${\boldsymbol \Phi}$ which is $N \times N$ \\
3: & Compute the eigenvalue decomposition of $\Delta\Delta^H$ and store the result in $\Lambda$ \\
4: & Generate a set of ${\boldsymbol \Phi}$ randomly with the power constraint $\rm{Tr}({\boldsymbol \Phi}_k{\boldsymbol \Phi}_k^\emph{H})\leq \rm{P_R}$ \\
5: & For all ${\boldsymbol \Phi}$, compute \\
& $\Theta(c)=\det\left({\boldsymbol I} + \frac{c}{2 \sqrt{2N_0}}{\boldsymbol \Phi}{\boldsymbol \Lambda}{\boldsymbol \Phi}^\emph{H} \right)^{-1}$ \\
6: & Choose the code matrix according to \\
&$\Theta_{opt}(c)=\arg\max_l \Theta_l(c)$\\
7: & Store the optimal code matrix ${\boldsymbol \Phi}_{opt}$ at the relay node \\
\hline
\end{tabular}
\vspace{-2em}
\end{table}

\section{Simulations}

The simulation results are provided in this section to assess the proposed scheme and algorithms. The cooperative MIMO system considered employs an AF protocol with the Alamouti STBC scheme in \cite{RC De Lamare} using QPSK modulation in a quasi-static block fading channel with AWGN. The effect of the direct link is also considered. It is possible to employ the DF protocol or use different number of antennas and relay nodes with a simple modification. The system is equipped with $n_r=1$ relay node and $N=2$ antennas at each node. In the simulations, we set the symbol power $\sigma^2_s$ as equal to 1, and the power of the adjustable code matrix in the ARMO algorithms are normalized. The $SNR$ in the simulations is the received $SNR$ which is calculated by $SNR = \frac{\parallel\sum_{k=1}^{n_r}{\boldsymbol \Phi}_{eq_k}[i]{\boldsymbol D}\parallel^2_F}{1+\parallel\sum_{k=1}^{n_r}{\boldsymbol \Phi}_{eq_k}[i]{\boldsymbol G}_{{eq}_k}[i]{\boldsymbol A}_{R_kD}[i]\parallel^2_F}$.

The upper bounds of the D-Alamouti, the R-Alamouti in \cite{Birsen Sirkeci-Mergen} and the adaptive Alamouti STC in C-ARMO RLS algorithm are shown in Fig. 2. The theoretical pairwise error probabilities provide the largest decoding errors of the three different coding schemes and as shown in the figure, by employing a randomized matrix at the relay node it decreases the decoding error upper bound. The bounds become tighter to the respective coding schemes as the SNR increases. The comparison of the simulation results in a better BER performance of the R-Alamouti and the D-Alamouti which indicates the advantage of using the randomized matrix at relay nodes. The C-ARMO RLS algorithm optimizes the randomized matrices after each transmission which contributes to a lower error probability upper bound, and the ML detection algorithm provides the optimal performance at the cost of a higher computation complexity.
\begin{figure}
\begin{center}
\def\epsfsize#1#2{0.825\columnwidth}
\epsfbox{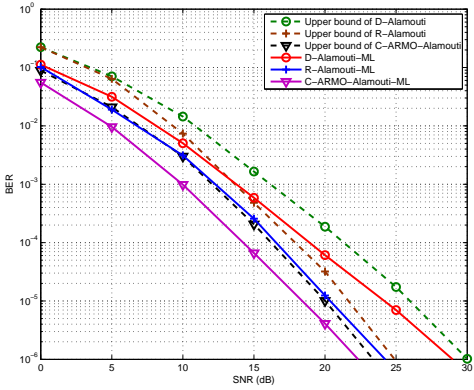}\vspace*{-1em} \caption{BER Performance vs.
$SNR$ for the Upper Bound of the Alamouti Schemes without the
Direct Link}\label{1}
\vspace{-3em}
\end{center}
\end{figure}

The proposed C-ARMO SG algorithm with a linear MMSE receiver is
compared with the SM scheme and the DSTC algorithms  in
\cite{YJing}, \cite{JHarshan}, \cite{Birsen Sirkeci-Mergen} and
\cite{Behrouz Maham} in Fig. 3.  It is worth to mention that the
coding schemes in the simulations are different. In the proposed
algorithm and the algorithms in \cite{YJing}, \cite{JHarshan} and
\cite{Birsen Sirkeci-Mergen}, a spatial multiplexing scheme is sent
from the source node and re-encoded at the relay node, while the
full-opportunistic code \cite{Behrouz Maham} requires an STC
encoding at the source node instead of re-encoding at the relay
node. The step sizes for the iterative optimization are $\beta=0.01$
and $\mu=0.03$, which are chosen according to \cite{Feuer}. The
results illustrate that without the direct link, by making use of
the STC technique, a significant performance improvement can be
achieved compared to the spatial multiplexing system. The RSTC
algorithm in \cite{Birsen Sirkeci-Mergen} outperforms the STC-AF
schemes in \cite{YJing} and \cite{JHarshan}, while the C-ARMO SG
algorithm can improve the performance by about $3$dB as compared to
the RSTC algorithm. The STC scheme in \cite{Behrouz Maham} achieves
a much better performance compared to other schemes although this
comparison must be considered with caution. In \cite{Behrouz Maham}
the standard $2 \times 2$ Alamouti STBC is employed at the source
node, which indicates the received matrix at the relay node is
amplified without the interference to the orthogonality of the code.
Moreover, encoding at the source node requires more time slots to
transmit so that the transmission rate is half compared to the
proposed C-ARMO algorithm. It is also worth to mention that the
C-ARMO algorithm can be employed in an opportunistic scheme to
achieve a better BER performance as both of the algorithms employ
the STCs and can perform the optimization at the destination node.
With the consideration of the direct link, the results indicate that
the diversity order can be increased, and using the C-ARMO SG
algorithm an improved performance is achieved with $2$dB of gain as
compared to employing the RSTC algorithm in \cite{Birsen
Sirkeci-Mergen} and $3$dB of gain as compared to employing the
traditional STC-AF algorithm in \cite{YJing}.
\begin{figure}
\begin{center}
\def\epsfsize#1#2{0.825\columnwidth}
\epsfbox{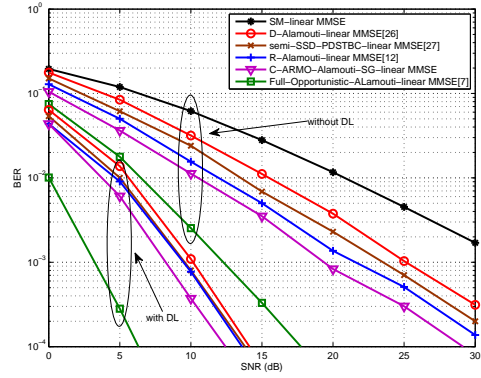}\vspace*{-1em} \caption{BER Performance vs.
$SNR$ for C-ARMO SG Algorithm with and without the Direct
Link}\label{2}
\vspace{-3em}
\end{center}
\end{figure}

In Fig. 4, BER curves of different Alamouti coding schemes and the proposed C-ARMO RLS algorithm with and without the direct link using an ML detector are compared. In Fig. 4, the R-Alamouti scheme improves the performance by about $4$dB without the direct link compared to the D-Alamouti scheme, and the C-ARMO RLS algorithm provides a significant improvement in terms of gains compared to the other DSTC schemes. When the direct link is considered, all the coding schemes can achieve the full diversity order and obtain lower BER performances compared to that without the direct link, and still the C-ARMO RLS algorithm which optimizes the adjustable code matrix achieves the lowest BER performance.
\begin{figure}
\begin{center}
\def\epsfsize#1#2{0.825\columnwidth}
\epsfbox{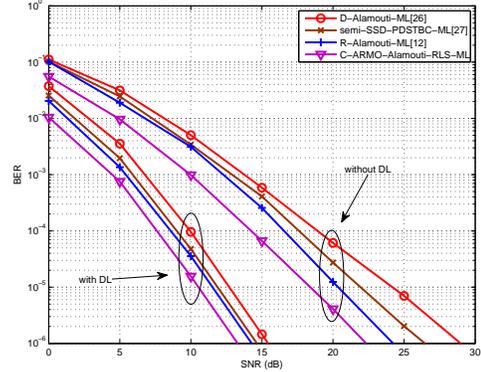}\vspace*{-1em} \caption{BER Performance vs.
$SNR$ for C-ARMO RLS Algorithm with and without the Direct
Link}\label{2}
\vspace{-3em}
\end{center}
\end{figure}

The simulation results shown in Fig. 5 illustrate the convergence property of the C-ARMO SG algorithm. All the schemes have an error probability of $1$ at the beginning, and after the first $20$ symbols are received and detected, the R-Alamouti scheme in \cite{YJing} achieves a better BER performance compared with the spatial multiplexing scheme and the R-Alamouti scheme in \cite{Birsen Sirkeci-Mergen} can reach a lower BER than the C-ARMO algorithm. With the number of received symbol increasing, the BER curve of the spatial multiplexing, the D-Alamouti and the R-Alamouti schemes are almost straight, while the BER performance of the C-ARMO algorithm can be further improved and obtain a fast convergence after receiving $140$ symbols.
\begin{figure}
\begin{center}
\def\epsfsize#1#2{0.825\columnwidth}
\epsfbox{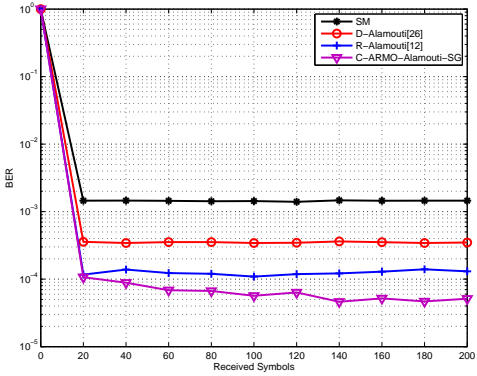}\vspace*{-1em} \caption{BER Performance vs.
Number of Samples for C-ARMO SG Algorithm without the Direct
Link}\label{3}
\vspace{-3em}
\end{center}
\end{figure}

The simulation results shown in Fig. 6 illustrate the influence of the feedback channel on the C-ARMO SG algorithm. As mentioned in Section I, the optimized code matrix will be sent back to each relay node through a feedback channel. The quantization and feedback errors are not considered in the simulation results in Fig. 3 and Fig. 4, so the optimized code matrix is perfectly known at the relay node after the C-ARMO algorithm; while in Fig. 6, it indicates that the performance of the proposed algorithm will be affected by the accuracy of the feedback information. In the simulation, we use $4$ bits to quantize the real part and the imaginary part of the element in the code matrix ${\boldsymbol \Phi}_{{eq_k}_j}[i]$, and the feedback channel is modeled as a binary symmetric channel with different error probabilities. As we can see from Fig. 6, by decreasing the error probabilities for the feedback channel with fixed quantization bits, the BER performance approaches the performance with the perfect feedback, and by making use of $4$ quantization bits for the real and imaginary part of each parameter in the code matrix, the performance of the C-ARMO SG algorithm is about 1dB worse with feedback error probability of $10^{-3}$.
\begin{figure}
\begin{center}
\def\epsfsize#1#2{0.825\columnwidth}
\epsfbox{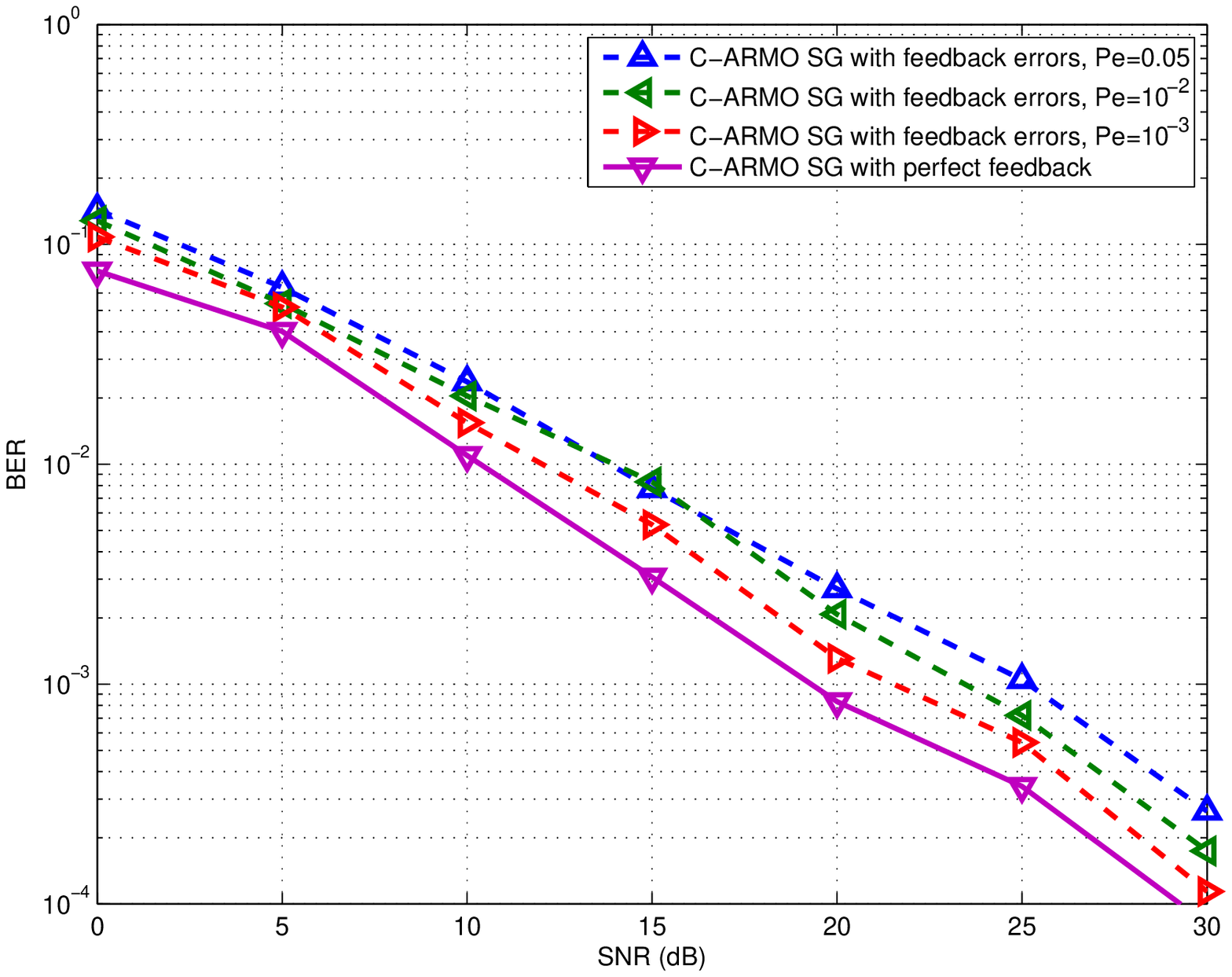}\vspace*{-1em} \caption{BER Performance vs. $SNR$ for C-ARMO Algorithm with Perfect and Imperfect Feedback Links, Quantization Bits = 4}\label{3}
\vspace{-3em}
\end{center}
\end{figure}

In Fig. 7, we plot the average error probability with respect to the SNR for the FD-ARMO algorithm and the C-ARMO SG algorithm with perfect feedback. The C-ARMO curve and the FD-ARMO curve outperforms the others because they optimize the adjustable code matrices with the same criterion, but $1$dB of gain has been obtained by the C-ARMO SG algorithm because the exact adjustable code matrix is transmitted back to the relay node in a delay-free and error-free feedback channel. While the FD-ARMO chooses the optimal adjustable code matrix by using the statistical information of the channel before transmission so that the performance will be influenced, resulting in a gain less than $1$dB.
\begin{figure}
\begin{center}
\def\epsfsize#1#2{0.825\columnwidth}
\epsfbox{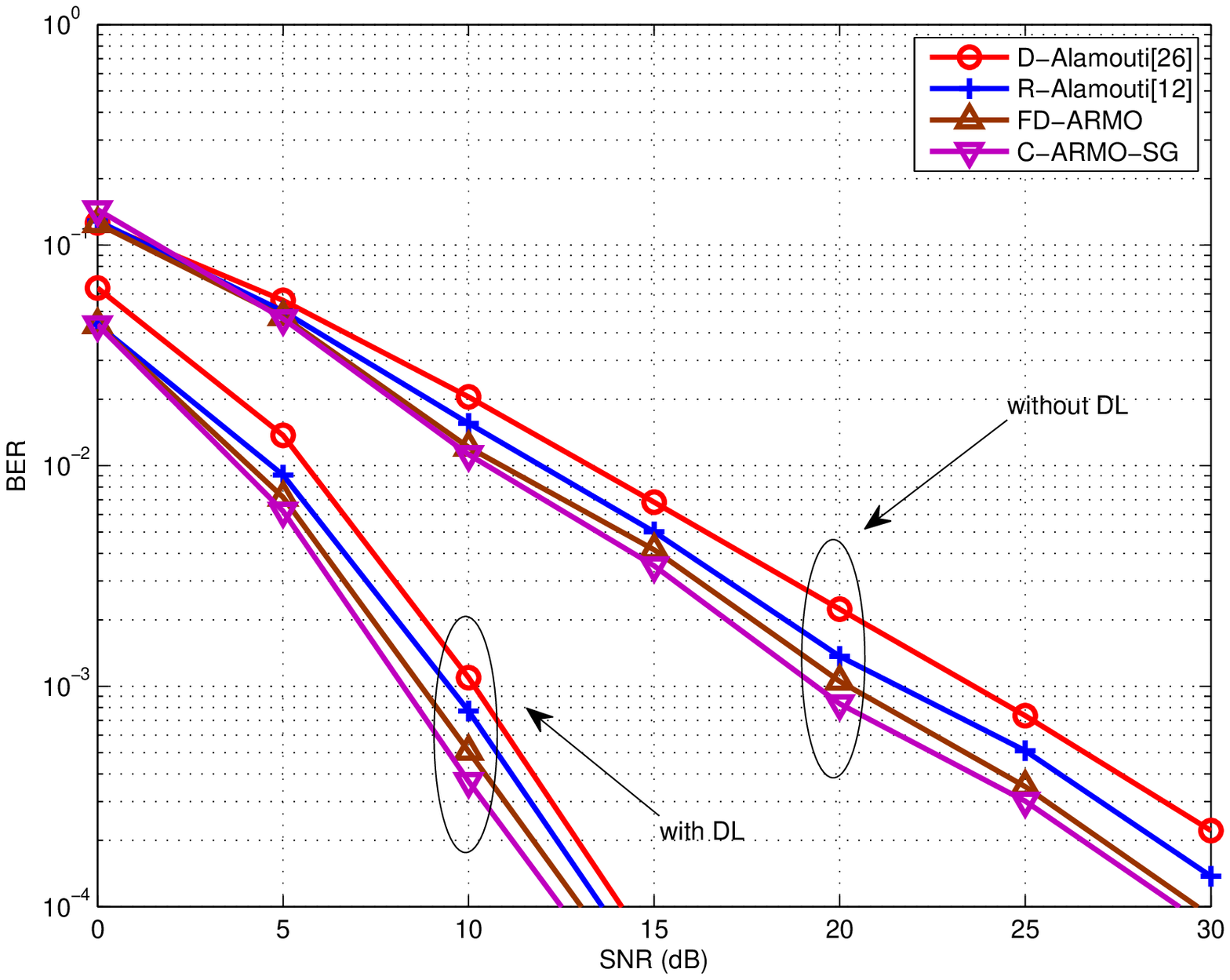}\vspace*{-1em} \caption{Full-Distributed ARMO Algorithm and C-ARMO SG Algorithm}\label{3}
\vspace{-3em}
\end{center}
\end{figure}

\section{Conclusion}

We have proposed centralized adaptive robust matrix optimization (C-ARMO) algorithms for the AF cooperative MIMO system using a linear MMSE receive filter and an ML receiver at the destination node. The pairwise error probability of introducing the adaptive DSTC in a cooperative MIMO network with the AF protocol has been derived. In order to eliminate the need for a feedback channel we have derived a fully-distributed ARMO (FD-ARMO) algorithm which can achieve a similar coding gain without the feedback as compared to the C-ARMO algorithms. The simulation results illustrate the advantage of the proposed ARMO algorithms by comparing them with the cooperative network employing the traditional DSTC scheme and the RSTC scheme. The proposed algorithms can be used with different DSTC schemes using the AF strategy and can also be extended to the DF cooperation protocol.

\appendices
\section{}
We show how to obtain the expression of the linear MMSE receive filter ${\boldsymbol w}_j[i]$ and the adjustable code matrix ${\boldsymbol \Phi}_{{eq_k}_j}[i]$ in equation (\ref{3.2}) and (\ref{3.3}) in Section III in the following.

The MSE optimization problem is given by
\begin{equation*}
    [{\boldsymbol w}_j[i],{\boldsymbol \Phi}_{eq_k}[i]] = \arg\min_{{\boldsymbol w}_j[i], {\boldsymbol \Phi}_{eq_k}[i]} E\left[\|s_j[i]-{\boldsymbol w}_j^\emph{H}[i]{\boldsymbol r}[i]\|^2\right], ~s.t.~
    \rm{Tr}({\boldsymbol \Phi}_{eq_k}[i]{\boldsymbol \Phi}_{eq_k}^\emph{H}[i])\leq \rm{P_R}.
\end{equation*}
We define a cost function associated with the optimization problem above and expand it as follows
\begin{equation}\label{a1}
\begin{aligned}
    \mathscr{L} &= E \left[ \|s_j[i]-{\boldsymbol w}_j^\emph{H}[i]{\boldsymbol r}[i] \|^2 \right]+\lambda(\rm{Tr}({\boldsymbol \Phi}_{eq_k}[i]{\boldsymbol \Phi}_{eq_k}^\emph{H}[i])-\rm{P_R})\\
    &= E \left[ s_j[i]s^*_j[i]\right] - {\boldsymbol w}_j^\emph{H}[i]E \left[{\boldsymbol r}[i]s^*_j[i]\right] - E \left[s_j[i]{\boldsymbol r}^\emph{H}[i]\right]{\boldsymbol w}_j[i] + {\boldsymbol w}_j^\emph{H}[i]E \left[{\boldsymbol r}[i]{\boldsymbol r}^\emph{H}[i]\right]{\boldsymbol w}_j[i]\\
    &~~~+\lambda(\rm{Tr}({\boldsymbol \Phi}_{eq_k}[i]{\boldsymbol \Phi}_{eq_k}^\emph{H}[i])-\rm{P_R}),
\end{aligned}
\end{equation}
where $\lambda$ stands for the Lagrange multiplier and should be determined before the calculation. It is worth to notice that the first, the third and the fifth terms are not functions of ${\boldsymbol w}^\emph{H}_j[i]$, so by taking the gradient of $\mathscr{L}$ with respect to ${\boldsymbol w}^\emph{H}_j[i]$ and equating the terms to $0$, we can obtain
\begin{equation}\label{a2}
    \mathscr{L}'_{{\boldsymbol w}^*_j[i]}= -E\left[{\boldsymbol r}[i]s^*_j[i]\right]+E \left[{\boldsymbol r}[i]{\boldsymbol r}^\emph{H}[i]\right]{\boldsymbol w}_j[i]=0.
\end{equation}
By moving the first term in (\ref{a2}) to the right-hand side and by multiplying the inverse of the auto-correlation of the received symbol vector, we obtain the expression of the linear MMSE receive filter as ${\boldsymbol w}_j[i] = {\boldsymbol R}^{-1}{\boldsymbol p}$, where the auto-correlation matrix ${\boldsymbol R}=E\left[{\boldsymbol r}[i]{\boldsymbol r}^\emph{H}[i]\right]$ and the cross-correlation vector ${\boldsymbol p}=E\left[{\boldsymbol r}[i]s_j^*[i]\right]$.

In order to obtain the expression of the adjustable code matrix ${\boldsymbol \Phi}_{eq_{k_j}}[i]$ we have to rewrite the received symbol vector ${\boldsymbol r}[i]$ as
\begin{equation}\label{a3}
    {\boldsymbol r}[i] = \sum_{k=1}^{n_r}{\boldsymbol \Phi}_{eq_k}[i]{\boldsymbol G}_{{eq}_k}[i]\tilde{{\boldsymbol s}}_{SR_k}[i] +{\boldsymbol n}_{RD}[i] = \sum_{k=1}^{n_r}\sum_{j=1}^{N}{\boldsymbol \Phi}_{eq_{k_j}}[i]{\boldsymbol g}_{eq_{k_j}}[i]\tilde{s}_{SR_{k_j}}[i]+{\boldsymbol n}_{RD}[i],
\end{equation}
where ${\boldsymbol \Phi}_{eq_{k_j}}[i]$ denotes the adjustable code matrix assigned to the $j$th received symbol $\tilde{s}_{SR_{k_j}}[i]$ at the $k$th relay node, and ${\boldsymbol g}_{eq_{k_j}}[i]$ stands for the $j$th column of the equivalent channel matrix ${\boldsymbol G}_{{eq}_k}[i]$. By substituting (\ref{a3}) into (\ref{a1}), the expression of $\mathscr{L}$ can be written as
\begin{equation*}
\begin{aligned}
    \mathscr{L} = & E \left[s_j[i]s^*_j[i]\right] - {\boldsymbol w}_j^\emph{H}[i]E [(\tilde{{\boldsymbol r}}+{\boldsymbol n}_{RD}[i])s^*_j[i]] - E [s_j[i]({\boldsymbol w}^\emph{H}_j[i](\tilde{{\boldsymbol r}}+{\boldsymbol n}_{RD}[i]))^\emph{H}] \\
    & + E [({\boldsymbol w}_j^\emph{H}[i](\tilde{{\boldsymbol r}}+{\boldsymbol n}_{RD}[i]))^\emph{H}{\boldsymbol w}^\emph{H}_j[i](\tilde{{\boldsymbol r}}+{\boldsymbol n}_{RD}[i])] +\lambda(\rm{Tr}({\boldsymbol \Phi}_{eq_k}[i]{\boldsymbol \Phi}_{eq_k}^\emph{H}[i])-\rm{P_R}),
\end{aligned}
\end{equation*}
where $\tilde{{\boldsymbol r}}=\sum_{k=1}^{n_r}\sum_{j=1}^{N}{\boldsymbol \Phi}_{eq_{k_j}}[i]{\boldsymbol g}_{eq_{k_j}}[i]\tilde{s}_{SR_{k_j}}[i]$. We do not have to consider the first and the second terms because they are not functions of ${\boldsymbol \Phi}_{eq_{k_j}}^\emph{H}[i]$ so taking the gradient of $\mathscr{L}$ with respect to ${\boldsymbol \Phi}_{eq_{k_j}}^*[i]$ these terms will disappear. The last three terms contain the sum of the adjustable code matrices, and we focus on the exact $j$th code matrix we need and consider the rest of the sum terms as constants. We can rewrite $\mathscr{L}$ as
\begin{equation}\label{a4}
\begin{aligned}
    \mathscr{L} = & -E \left[s_j[i]({\boldsymbol w}^\emph{H}_j[i]{\boldsymbol \Phi}_{eq_{k_j}}[i]{\boldsymbol g}_{eq_{k_j}}\tilde{s}_{SR_{k_j}}[i])^\emph{H}\right]+\lambda({\boldsymbol \Phi}_{eq_{k_j}}[i]{\boldsymbol \Phi}^\emph{H}_{eq_{k_j}}[i]-\rm{P_R}{\boldsymbol I}) \\ & + E [({\boldsymbol w}_j^\emph{H}[i]{\boldsymbol \Phi}_{eq_{k_j}}[i]{\boldsymbol g}_{eq_{k_j}}[i]\tilde{s}_{SR_{k_j}}[i])^\emph{H}{\boldsymbol w}^\emph{H}_j[i]{\boldsymbol \Phi}_{eq_{k_j}}[i]{\boldsymbol g}_{eq_{k_j}}[i]\tilde{s}_{SR_{k_j}}[i]],\\
\end{aligned}
\end{equation}
and by taking the gradient of $\mathscr{L}$ in (\ref{a4}) with respect to ${\boldsymbol \Phi}^*_{eq_{k_j}}[i]$ and equating the terms to zero, we can obtain ${\boldsymbol \Phi}_{eq_{k_j}}[i] = \tilde{\boldsymbol R}^{-1}\tilde{\boldsymbol P}$, where $\tilde{\boldsymbol R}=E\left[s_j[i]\tilde{s}_{SR_{k_j}}[i]{\boldsymbol w}_j[i]{\boldsymbol w}^\emph{H}_j[i]+\lambda{\boldsymbol I}\right]$ and $\tilde{\boldsymbol P}=E\left[s_j[i]\tilde{s}_{SR_{k_j}}[i]{\boldsymbol w}_j[i]{\boldsymbol g}^\emph{H}_{eq_{k_j}}[i]\right].
$
\section{}
We show the detailed derivation of the C-ARMO SG algorithm in this section. First, we have to rewrite the received symbol vector ${\boldsymbol r}_{R_kD}$ transmitted from the $k$th relay node. By employing the AF cooperative strategy and space-time coding schemes at the relay node, the received symbol vector at the relay nodes will be amplified and re-encoded prior to being forwarded to the destination node. Let us first define the amplified symbol vector before re-encoding as
\begin{equation}\label{b1}
\begin{aligned}
    \tilde{{\boldsymbol s}}_{SR_k}[i]& ={\boldsymbol A}_{R_kD}[i]({\boldsymbol F}_{SR_k}[i]{\boldsymbol s}[i] + {\boldsymbol n}_{SR_k}[i]) = {\boldsymbol A}_{R_kD}[i]{\boldsymbol F}_{SR_k}[i]{\boldsymbol s}[i] + {\boldsymbol A}_{R_kD}[i]{\boldsymbol n}_{SR_k}[i]\\
    & = {\boldsymbol F}_{R_k}[i]{\boldsymbol s}[i]+{\boldsymbol n}_{R_k}[i],
\end{aligned}
\end{equation}
where ${\boldsymbol A}_{R_kD}[i]$ denotes the $N \times N$ amplify matrix at the $k$th relay node. The symbol vector $\tilde{{\boldsymbol s}}_{SR_k}[i]$ will be mapped to an $N \times T$ space-time code matrix ${\boldsymbol M}(\tilde{{\boldsymbol s}})$, and multiplied by an adjustable code matrix which is generated randomly before being forwarded to the destination node. By substituting (\ref{b1}) into (\ref{2.4}), the relationship between all the relay nodes and the destination node can be written as
\begin{equation}\label{b2}
\begin{aligned}
    {\boldsymbol r}_{RD} &= \sum_{k=1}^{n_r}{\boldsymbol \Phi}_{eq_k}[i]{\boldsymbol G}_{{eq}_k}[i]({\boldsymbol F}_{R_k}[i]{\boldsymbol s}[i]+{\boldsymbol n}_{R_k}[i])+{\boldsymbol n}_{RD}[i] = \sum_{k=1}^{n_r}{\boldsymbol \Phi}_{eq_k}[i]{\boldsymbol D}_k[i]{\boldsymbol s}[i]+{\boldsymbol n}_D[i]\\
    &=\sum_{k=1}^{n_r}\sum_{j=1}^{N}{\boldsymbol \Phi}_{eq_{k_j}}[i]{\boldsymbol d}_{k_j}[i]s_j[i]+{\boldsymbol n}_D[i],
\end{aligned}
\end{equation}
where the $NT \times N$ matrix ${\boldsymbol D}_k[i]$ contains all the channel information between the source node and the $k$th relay node, and between the $k$th relay node and the destination node. The noise vector at the destination node ${\boldsymbol n}_D[i]$ is Gaussian with covariance matrix $\sigma^2(1+Tr(\sum_{k=1}^{n_r}{\boldsymbol \Phi}_{eq_k}[i]{\boldsymbol D}_k[i])){\boldsymbol I}_N$. By substituting (\ref{b2}) into (\ref{3.1}), we can rewrite the MSE optimization problem as
\begin{equation}\label{b3}
\begin{aligned}
    &[{\boldsymbol w}_j[i],{\boldsymbol \Phi}_{eq_{k_j}}[i]] =
     \arg\min_{{\boldsymbol w}_j[i], {\boldsymbol \Phi}_{eq_{k_j}}[i]} E\left[\|s_j[i]-{\boldsymbol w}_j^\emph{H}[i](\sum_{k=1}^{n_r}\sum_{j=1}^{N}{\boldsymbol \Phi}_{eq_{k_j}}[i]{\boldsymbol d}_{k_j}[i]s_j[i]+{\boldsymbol n}_D[i])\|^2\right],\\
    & ~~~~~~~~~~~~~~~~~~~~~~~~~~~~~~~~~~~~s.t.~~ \rm{Tr}(\sum_{j=1}^{N}{\boldsymbol \Phi}_{{eq_k}_j}[i]{\boldsymbol \Phi}_{{eq_k}_j}^\emph{H}[i])\leq \rm{P_R}.
\end{aligned}
\end{equation}

By taking the instantaneous gradient of $\mathscr{L}$ in (\ref{a1}) with respect to ${\boldsymbol w}^\emph{H}_j[i]$ and ${\boldsymbol \Phi}^\emph{H}_{eq_{k_j}}[i]$ we can obtain
\begin{equation}\label{b4}
\begin{aligned}
     \nabla \mathscr{L}_{{\boldsymbol w}_j^*[i]} & = \nabla E\left[\|s_j[i]-{\boldsymbol w}_j^\emph{H}[i]{\boldsymbol r}[i]\|^2\right]_{{\boldsymbol w}_j^*[i]} = (s_j[i]-{\boldsymbol w}_j^\emph{H}[i]{\boldsymbol r}[i])^\emph{H}\nabla_{{\boldsymbol w}_j^*[i]}(s_j[i]-{\boldsymbol w}_j^\emph{H}[i]{\boldsymbol r}[i])\\
     & = -e_j^*[i]{\boldsymbol r}[i],\\
     \nabla \mathscr{L}_{{\boldsymbol \Phi}_{{eq_k}_j}^*[i]} &= \nabla E\left[\|s_j[i]-{\boldsymbol w}_j^\emph{H}[i](\sum_{k=1}^{n_r}\sum_{j=1}^{N}{\boldsymbol \Phi}_{eq_{k_j}}[i]{\boldsymbol d}_{k_j}[i]s_j[i]+{\boldsymbol n}_{RD}[i])\|^2\right]_{{{\boldsymbol \Phi}^*_{{eq_k}_j}}[i]} \\ & = \nabla_{{{\boldsymbol \Phi}^*_{{eq_k}_j}}[i]}(s_j[i]-{\boldsymbol w}_j^\emph{H}[i](\sum_{k=1}^{n_r}\sum_{j=1}^{N}{\boldsymbol \Phi}_{eq_{k_j}}[i]{\boldsymbol d}_{k_j}[i]s_j[i]+{\boldsymbol n}_{RD}[i]))^\emph{H}(s_j[i]-{\boldsymbol w}_j^\emph{H}[i]{\boldsymbol r}[i]) \\
    &= -e_j[i]s^*_j[i]{\boldsymbol w}_j[i]{\boldsymbol d}_{k_j}^\emph{H}[i],
\end{aligned}
\end{equation}
where $e_j[i]=s_j[i]-{\boldsymbol w}_j^\emph{H}[i]{\boldsymbol r}[i]$ stands for the $j$th detected error. By employing step sizes $\beta$ and $\mu$ for the receive filter and the code matrix recursions, respectively, we obtain the C-ARMO SG algorithm derived as
\begin{equation*}
\begin{aligned}
     {\boldsymbol w}_j[i+1] & = {\boldsymbol w}_j[i] + \beta (e_j^*[i]{\boldsymbol r}[i]),\\
     {\boldsymbol \Phi}_{{eq_k}_j}[i+1] &={\boldsymbol \Phi}_{{eq_k}_j}[i] +\mu (e_j[i]s^*_j[i]{\boldsymbol w}_j[i]{\boldsymbol d}_{k_j}^\emph{H}[i]).
\end{aligned}
\end{equation*}

\bibliographystyle{IEEEtran}

\end{document}